\documentclass[final]{ring} 

\usepackage[lofdepth,lotdepth]{subfig} 
\usepackage{amssymb}
\usepackage{bbold}
\usepackage{color}
\usepackage{multirow}
\usepackage{float}

\newcommand{\1}{\mathbb{1}}

\newcommand{\RR}{\mathbb{R}}
\newcommand{\NN}{\mathbb{N}}

\newcommand{\cs}{\mathbf{s}}
\newcommand{\xx}{\mathbf{x}}

\newcommand{\dd}{\mathbf{d}}
\graphicspath{%
{./Figures/}%
}

\title{Hug model: parameter estimation via the ABC Shadow algorithm}

\author[1]{Christophe Reype}
\author[1]{Radu S. Stoica}
\author[2]{Didier Gemmerlé}
\author[3]{Antonin Richard}
\author[4]{Madalina Deaconu}

\affil[1]{Universit\'e de Lorraine, CNRS, Inria, IECL, F-54000 Nancy, France}
\affil[2]{Universit\'e de Lorraine, IECL, F-54000 Nancy, France}
\affil[3]{Universit\'e de Lorraine, CNRS, GeoRessources, F-54000 Nancy, France}
\affil[4]{Universit\'e de Lorraine, Inria, IECL, F-54000 Nancy, France}
\shortauthor{Reype et al.} % 
\shorttitle{Robustness of the Hug model}

\begin{document}
\maketitle

\begin{abstract}

%\RS{
Studying geological fluids mixing systems allows to understand interaction among water sources. The Hug model is an interaction point process model that can be used to estimate the number and the chemical composition of the water sources involved in a geological fluids mixing system from the chemical composition of samples~\cite{Rey20,Rey22,reype2022probabilistic}. The source detection using the Hug model needs prior knowledge for the model parameters. The present work shows how the parameter estimation method known as the ABC Shadow algorithm~\cite{StoPhi17,StoDea21} can be used in order to construct priors for the parameters of the Hug model. The long term perspective of this work is to integrate geological expertise within fully unsupervised models.%}

\end{abstract}

\section*{Introduction}

The building of conceptual and quantitative models of fluid and mass transfer in the sub-surface and the Earth's crust rely on the analysis of hydrochemical data \cite{Fau97,YarBod14,IngSan06}. Groundwater is the result of a mixing between water sources. Hence, the chemical composition of a sample of groundwater is link to the chemical composition of the water sources. The mass transfer occurring in the sub-surface is deduced by estimating the contribution of the water sources in samples. 

\begin{figure}[H]
\centering
\includegraphics[scale=0.5]{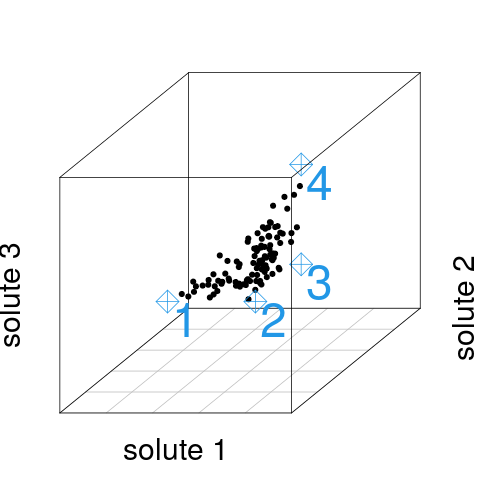}
\caption{Mixing scenario of four sources (in blue) in the three-dimensional space of hydrochemical parameters (solute1, solute2, solute3) called data space. The dots represent the position of the sample in the data space.}
\label{fig: mixing}
\end{figure}

However, the number and the composition of the water sources are not known. Thus, the water sources should be detected: their number and their chemical composition should be estimated. A toy example of mixing scenario between four sources (in blue) resulting in $200$ samples (dots) is shown in Figure \ref{fig: mixing}. This mixing system consider the composition in three hydrochemical parameters that can be concentration of ions or molecules, isotopic composition or ratios of hydrochemical parameters. The space made by the composition in the hydrochemical parameters is called data space.

%\RS{
In the literature, source detection is done mostly either graphically or in a supervised statistical analysis. Moreover, the detection often considers a subset of the hydrochemical parameters, that is to say the detection is performed by artificially combining results obtained on projections of the entire data space. Such a final result lacks of mathematical rigor and it can be biased.  For instance, consider the situation when two sources have the same position in the projected data space but a different position in the whole data space. In this situation, the exact number of sources is never precisely detected.%}

%In the literature, source detection is done mostly either graphically or in a supervised statistical analysis. Moreover, the detection often considers a subset of the hydrochemical parameters, that is to say the detection is performed on projections of the data space. This result is biased due to projection if, for example, two sources have the same position in the projected data space but a different position in the whole data space. Hence, the real sources may not be detected. Source detection rely on the user: the number of sources should be known for the End-Member Mixing Analysis \cite{Wel97} and other geometrical method, that estimate the sources as the vertices of the smallest convex hull containing the data \cite{Pint20}. 

The Hug model~\cite{Rey20,Rey22,reype2022probabilistic} is a point process especially built to detect the number of sources and their chemical composition in hydrochemical data. This point process considers the data as a cloud of points in a multidimensional space, with coordinates being the value of each hydrochemical parameter. The sources distribution in the data space is governed by a Gibbs point process. The density probability of this Gibbs point process is built by taking into account knowledge on the analyzed mixing system and assumptions used in the graphical detection methods. %The first assumption is to minimize the number of sources. The second assumption is to select sources that explain the data (\textit{i.e.} the convex hull of the sources tends to enclose the data). The third assumption is to consider that the data are representative of the mixing system (\textit{i.e.} the convex hull of the sources is outlined by the data). The last assumption is to consider that the composition of the sources are significantly different from another.
The conditions for using the Hug model are as follows. The datasets (\textit{i.e.} the samples) are the result of a conservative mixing (\textit{i.e.} no chemical reaction affects the considered hydrochemical parameters during the mixing process). The composition of the sources are supposed the same, or at least not significantly different for each data point.% The hydrochemical parameters considered should induce a linear mixing, that is to say that the data points are described as in equation \ref{eq: mixing}, and not any curvature in the mixing trends \cite{LanVoc78}.

The sources are estimated by the configuration that maximizes the joint probability density controlling the sources and the parameters distributions. This probability density is simulated by Markov chains Monte Carlo methods. The optimization procedure is implemented via a simulated annealing procedure, hence avoiding local maximum. 
%The sources are updated by projecting the data on the two-dimensional spaces made by selecting a paired set of two hydrochemical parameters. On each of these projected data space, called plane, the detection is an automation of the graphical detection. This update is made possible by using the Metropolis-Hastings within Gibbs sampler. The optimization procedure is implemented via a simulated annealing procedure, hence avoiding local maximum. 

The Hug model need model parameters set by the user to detect the sources. This paper presents a method to estimate the model parameters from bidimensional sources in order to improve the detected sources: the ABC Shadow algorithm \cite{StoPhi17,StoDea21}.

%This paper has two goals. The first goal is to test the robustness of the Hug model. More specifically, this paper will test the sensibility of the Hug model to the data uncertainties. The second goal is to estimate the model parameters in order to transform the Hug model into an unsupervised model.

\section{Materials and methods}
\label{sec:Method}

\subsection{Hug model}
%The considered data set is 
Considered data sets are made of measures of hydrochemical parameters for the collected samples.
Each sample is a point in an abstract multidimensional space, called data space. Each sample coordinate represents the composition of a given hydrochemical parameters. The observed samples are the result of the mixing between the unknown sources.

%The data considered are measures of hydrochemical parameters of different samples. The data are seen as spatial data: each sample is a point in the abstract space, called data space, which each coordinate is the composition in a given hydrochemical parameters. The samples are the results of the mixing between the sources.

The Hug model is a Gibbs point process that can be used to detect the number and the composition of the sources involved in a mixing system from the data. This point process takes into account the following assumptions:
\begin{itemize}
    \item[(a)] the data points are rather close to the sources,
    \item[(b)] the data points are barycenters of the sources,
    \item[(c)] the number of sources is minimized,
    \item[(d)] the composition of the sources are significantly different from one source to another.
\end{itemize}

These assumptions are “translated” into a function $U$ called energy function \cite{Rey22}. Assume the model parameters vector $\theta$ known, the probability density of the Gibbs point process representing the Hug model writes as
\begin{equation}
    p_\dd(\cs|\theta)=\frac{exp(-U(\cs|\theta))}{Z(\theta)},
\end{equation}
with $Z(\theta)$ the normalizing constant.

This probability density is simulated by using Metropolis-Hastings dynamics~\cite{MollWaag04,Lies00}. The detection of the sources is made by estimating the configuration of sources that maximizes the previous probability density. This optimization problem is solved by using a simulated annealing algorithm that applies the previous Metropolis-Hastings dynamics~\cite{Rey20,Rey22,reype2022probabilistic}. The Hug model is implemented by adapting the C++ library DRLib \cite{GemSto22}. The purpose of this library is the modelling, the simulation and the inference of marked point processes.

\subsection{Parameters estimation}
The source detection depends on the values of the model parameters. Previous cited work used Approximate Bayesian Computation (ABC) strategy to chose appropriate values for the model parameters~\cite{Blu10}. The principle of this technique is to compare realizations of simulated patterns of sources with configurations of known sources $\cs_{obs}$. The parameters that produce statistics close to the observation are kept for building the posterior. This scheme is presented below:

\fbox{
\begin{minipage}{0.95\linewidth}
\textbf{ABC algorithm:} set $\epsilon\in\RR_+$.
\begin{enumerate}
\item[1)] generate $\psi$ by $p(\psi)$ 
\item[2)] generate $\xx$ by $p(\xx|\psi)$
\item[3)] while $d\left(t(\cs_{obs}),t(\xx)\right) > \epsilon$
\item[4)] accept $\psi$ as parameter of interest 
\end{enumerate}
\end{minipage}
}

The ABC method should choose a tolerance threshold $\epsilon$, a distance $d$ between the observed statistics of the pattern of  known sources $t(\cs_{obs})$ and the simulated statistics obtained from the model. An important problem to be considered is that for numerical efficiency, the proposed $\phi$ should be close to the parameter of interest $\theta$. But this is exactly the problem to be solved: the prior to be used should be close to the unknown posterior.

In this paper, the ABC Shadow, \cite{StoPhi17,StoDea21} is used to estimate the model parameters from the posterior approximation. This choice is adopted since this algorithm does not exhibit the problems mentioned before. The ABC Shadow algorithm steps are:

\fbox{
\begin{minipage}{0.95\linewidth}
\textbf{ABC Shadow algorithm.} Data: $\cs_{obs}$ the known sources, $\theta^{(0)} \in\Theta$ the parameter initial value, $\Delta\in\RR_+$ the perturbation parameter and $N\in\NN$ the number of iterations.
\begin{enumerate}
\item[1)] generate $\xx$ by $p(\xx|\theta^{(0)})$ 
\item[2)] For $k=1,\dots,N$ \begin{itemize}
\item[a)] generate $\psi$ uniformly on the ball $b(\theta_k,\Delta/2)$
\item[b)] set $\theta^{(k)}=\psi$ with probability 
\begin{equation*}
    \alpha(\theta\longrightarrow\psi)=\min\left\{1,\frac{p(\psi|\cs_{obs})}{p(\theta|\cs_{obs})}\frac{Z(\psi)\exp(-U(\xx|\theta))\1_{b(\psi,\Delta/2)}(\theta)}{Z(\theta)\exp(-U(\xx|\psi))\1_{b(\theta,\Delta/2)}(\psi)}\right\}.
\end{equation*}else set and $\theta^{(k)}=\theta^{(k-1)}$ 
\end{itemize}
\item[3)] set $\theta=\theta^{(N)}$
\end{enumerate}
\end{minipage}
}

\section{Results}

\subsection{Parameter estimation}

%The parameter estimation is made on another synthetic data set resulting of a mixing of four sources in a bi-dimensional data. The real sources are known.

The real data set, considered here, is presented in \cite{Pint20}. In this data set the stable isotopic composition of chlorine $\delta^{37}$Cl, the stable isotopic composition of bromine $\delta^{81}$Br and the stable isotopic composition of helium Rc/Ra ($^3$He$/^4$He normalized to that of the Atmosphere and corrected for the air component) are measured on $m=75$ samples from geothermal wells from Mexico and are supposed to be the result of a three-source mixing system  (mantle, subduction and crust). Note that halogens and noble gases behave conservatively during fluid mixing and that the mixing trends in the two planes considered here are not affected by curvature. In \cite{Pint20}, a source detection method is presented: on bi-dimensional data, the sources are the vertices of the smallest triangle, in terms of area, that contains the data. These bi-dimensional sources are used to estimate the model parameters of the Hug model by the ABC Shadow algorithm.

\begin{figure}[H]
\centering 
\subfloat[first normalized plane ]{\includegraphics[scale=0.4]{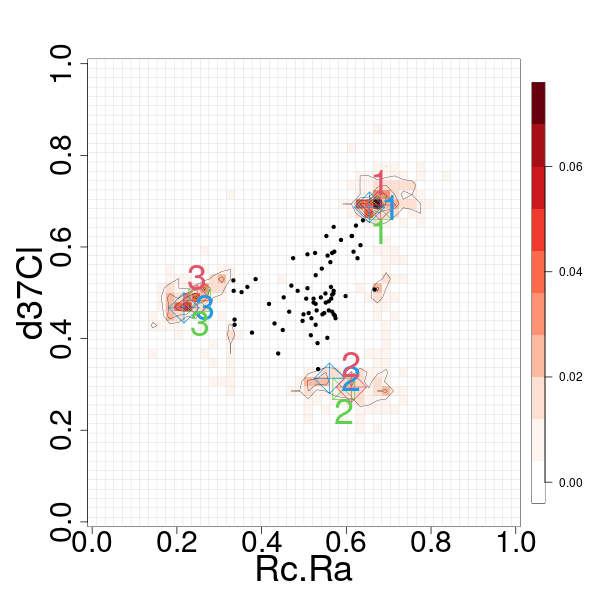}} 
\subfloat[second normalized plane ]{\includegraphics[scale=0.4]{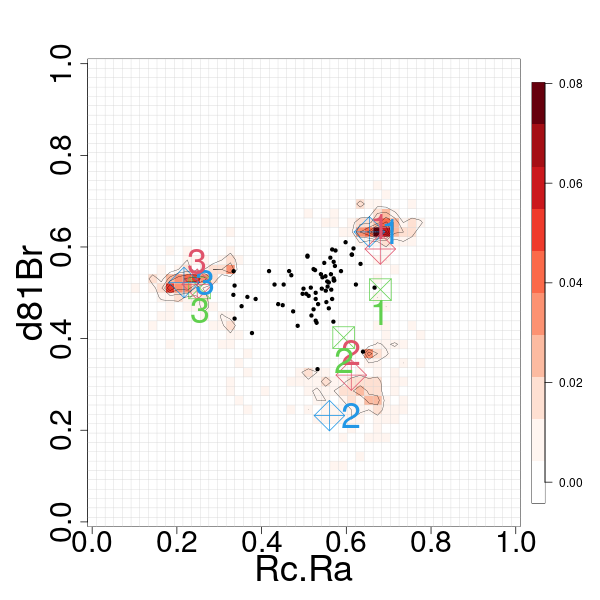}} 
\caption{
Visit map computed applied to the real data set~\cite{Pint20} using the empirical prior for $\theta$ . The blue symbols represent the sources reconstructed by \cite{Pint20}. The detection results obtained with the Hug model are represented by the green symbols and the red symbols. The first  ones represent the clusters centers (means) obtained after the aggregation of the outputs of the optimization algorithm, while the second ones indicate the median points of each cluster.
} 
\label{plot: level empirical Pinti} 
\end{figure}

%\RS{
The empirical $\theta$ selected in \cite{Rey22} gives the visit map of Figure \ref{plot: level empirical Pinti}. The visit map estimates the probability that a point in the data space is in contact or visited by a source. The detected sources are given in Table \ref{tab: proposed empirical pinti}. The distances between the detected sources and the known sources are the absolute values of the differences of their coordinates. They are given in Table \ref{tab: error empirical pinti}.
%}

%The empirical $\theta$ selected in \cite{Rey22} gives the visit map of Figure \ref{plot: level empirical Pinti}. The visit map is computed by counting the number of simulated sources in each cell of a grid. For each cell, the probability of having a source (\textit{e.g.} the number of sources in a cell divided by the number of considered simulations) is given. The proposed sources are given in Table \ref{tab: proposed empirical pinti}. The distance between the proposed sources and the known sources is the absolute difference between their coordinates and is given in Table \ref{tab: error empirical pinti}.

\begin{table}[H]
    \begin{tabular}{|c||c|c|c|c|c|c|c|c|c|}
    \hline
       \multirow{2}{*}{} & \multicolumn{2}{|c|}{Rc/Ra} & \multicolumn{2}{|c|}{$\delta^{37}$Cl} & \multicolumn{2}{|c|}{ $\delta^{81}$Br} \\
        \cline{2-7}
        & mean   & sd& mean  & sd& mean &  sd\\
        \hline\hline
1 (mantle) &0.69  &0.06 &0.7  &0.07 &0.59  &0.13\\
\hline
2 (subduction) &0.6& 0.08 &0.29& 0.07& 0.3&  0.14\\
\hline
3 (crust) &0.24 &0.07& 0.47& 0.06& 0.52 &0.06\\
\hline
    \end{tabular}   
    \caption{Proposed sources computed from the empirical $\theta$.}
    \label{tab: proposed empirical pinti}
\end{table}

\begin{table}[H]
    \begin{tabular}{|c||c|c|c||c|}
    \hline
        \textbf{Sources} & Rc/Ra &$\delta^{37}$Cl & $\delta^{81}$Br & Mean Distance by Source\\
        \hline
1 &0.01   &  0.00 &0.13 & 0.05\\
\hline
2 & 0.01  &   0.02 &0.17 & 0.07\\
\hline
3 & 0.03   &  0.02 &0.01 & 0.02\\
\hline\hline
Mean Distance by Dimension&0.02 &    0.01 &    0.10   &  0.04\\
        \hline
    \end{tabular}
    \caption{Distance between the proposed sources computed from the empirical $\theta$ and the sources estimated in \cite{Pint20}.}
\label{tab: error empirical pinti}
\end{table}

\begin{figure}[H]
\centering 
\subfloat[$\theta_1$]{\includegraphics[scale=0.3]{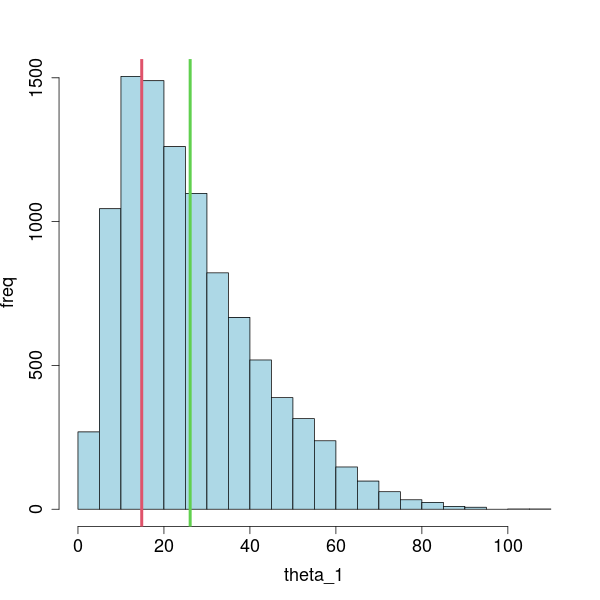}}
\subfloat[$\theta_2$]{\includegraphics[scale=0.3]{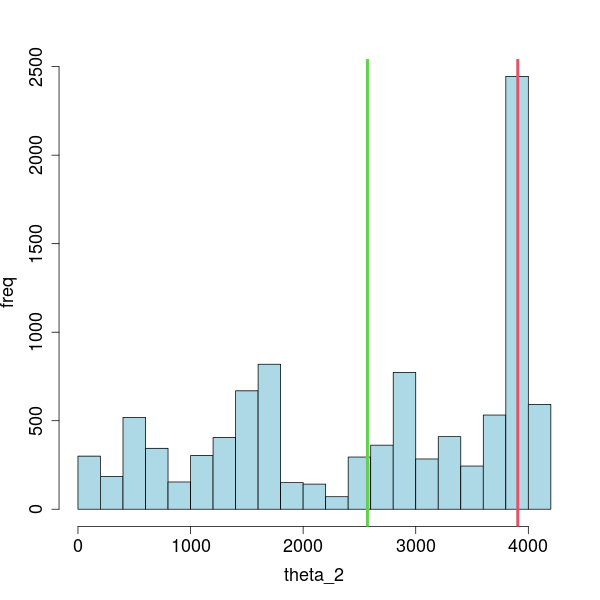}}\\
\subfloat[$\theta_3$]{\includegraphics[scale=0.3]{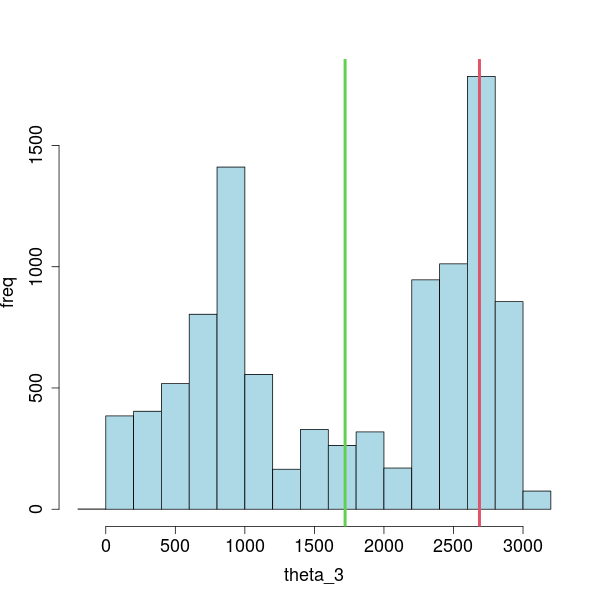}}
\subfloat[$\theta_4$]{\includegraphics[scale=0.3]{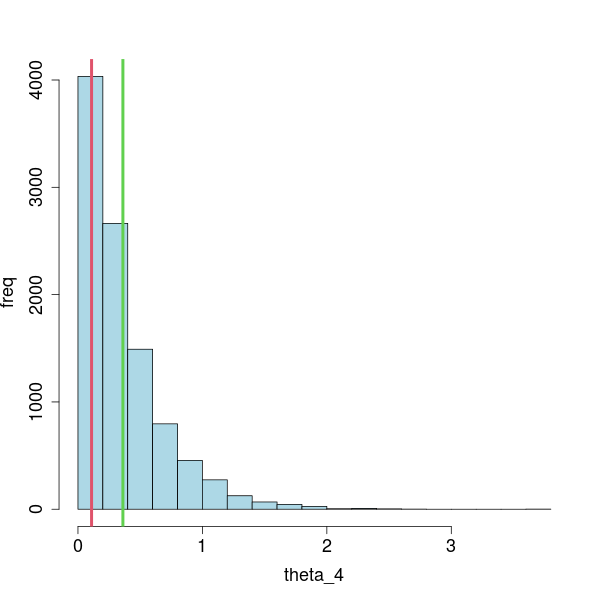}}
\caption{Histogram of the posterior approximation of $\theta$. The mean value are colored in green and the modal value in red.} 
\label{plot: abc hist} 
\end{figure}

For the ABC Shadow algorithm, the ranges of value that $\theta$ can takes are given in Table \ref{tab: ABC init}. The summary statistics of the %\RS{
posterior approximation%}
of $\theta$ are given in Table \ref{tab: ABC summary} %\RS{
while the histograms of $\theta$ are presented in Figure \ref{plot: abc hist}.%}

\begin{table}[H]
    \begin{tabular}{|c||c|c|c|}
    \hline
        & Initial value & min value & max value\\
        \hline
$\theta_1$ &1.0  &0.0 &10000.0 \\
\hline
$\theta_2$ & 1.0& 0.0 &10000.0 \\
\hline
$\theta_3$ &1.0 & -10000.0 &10000.0\\
\hline
$\theta_4$ &1.0 & 0.0 &10000.0\\
\hline
    \end{tabular}
    \caption{Initialization of the parameter for the ABC Shadow algorithm.}
\label{tab: ABC init}
\end{table}

\begin{table}[H]
    \begin{tabular}{|c||c|c|c|c|c|c|}
    \hline
&Q25 &Q50&Q75&mean&mode&sd\\
\hline
$\theta_1$&13.98& 22.77& 35.08 &26.1& 14.83& 15.9\\
\hline
$\theta_2$&1500.25 &2893.75 &3857.65& 2570.99 &3906.32 &1282.21\\
\hline
$\theta_3$&865.92& 1942.69& 2624.99& 1720.65& 2686.69& 932.69\\
\hline
$\theta_4$&0.13 &0.26& 0.49 &0.36& 0.11 &0.33\\
\hline
\hline
    \end{tabular}
    \caption{Summary statistics of the posterior estimation of $\theta$.}
\label{tab: ABC summary}
\end{table}

%Two candidates for $\theta$ can be used: the mean value of the estimation of $\theta$ and the modal value of the estimation of $\theta$. The results of the detection with each candidate are given in Figure \m{ABC detection}.
%If the sources are known, the ABC Shadow algorithm is able to give good estimators of the real parameters $\theta$.

\begin{figure}[H]
\centering 
\subfloat[first normalized plane]{\includegraphics[scale=0.25]{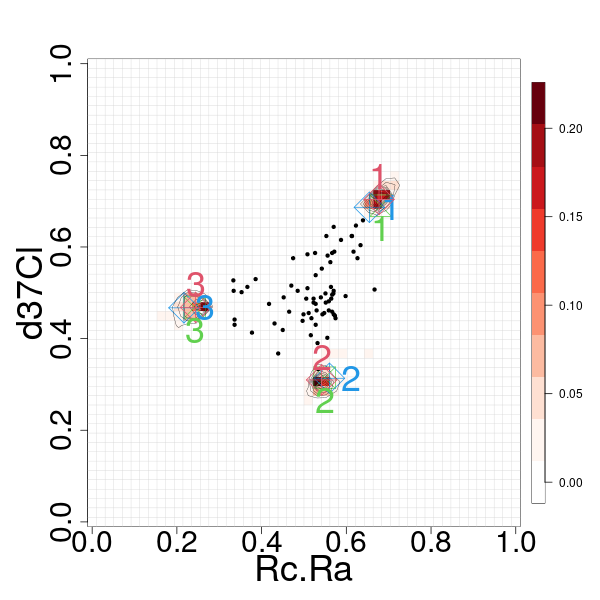}}
\subfloat[second normalized plane]{\includegraphics[scale=0.25]{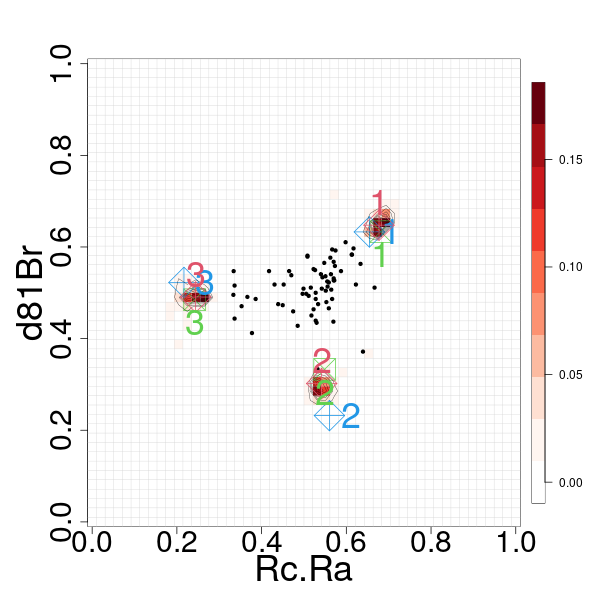}}
\caption{Visit map computed from the estimated $\theta$ for the real dataset~\cite{Pint20}. The blue symbols represent the sources reconstructed by \cite{Pint20}, the green symbols are the centers of the clusters and the red symbols are the median points.} 
\label{plot: level estimated Pinti} 
\end{figure}

The posterior mean values of $\theta$ are used to make a new detection of the sources by the Hug model in Figure \ref{plot: level estimated Pinti}. Graphically, the areas containing the sources are smaller than with the empirical $\theta$. The proposed sources computed from the estimation of $\theta$ are given in Table \ref{tab: proposed estimated pinti}. As expected, the standard deviations of the sources are smaller with the estimated $\theta$. Moreover, the distance between the proposed sources and the sources presented in \cite{Pint20} are also smaller. %\RS{
Based on this analysis, the detection method performs more accurately whenever the estimated $\theta$ are used.%}

    \begin{table}[H]
    \begin{tabular}{|c||c|c|c|c|c|c|c|c|c|}
    \hline
       \multirow{2}{*}{} & \multicolumn{2}{|c|}{``Rc/Ra''} & \multicolumn{2}{|c|}{``delta37Cl''} & \multicolumn{2}{|c|}{ ``delta81Br''} \\
        \cline{2-7}
        & mean   & sd& mean  & sd& mean &  sd\\
        \hline\hline
1 (mantle) &0.68  & 0.04 & 0.69  & 0.08 & 0.63  & 0.08\\
\hline
2 (subduction)  &0.55  & 0.03 & 0.31  & 0.03 & 0.33  & 0.1\\
\hline
3 (crust) &0.24  & 0.04 & 0.47  & 0.03 & 0.48  & 0.04\\
\hline
    \end{tabular} 
    \caption{Proposed sources computed from the estimated $\theta$.}
    \label{tab: proposed estimated pinti}
\end{table}

\begin{table}[H]
    \begin{tabular}{|c||c|c|c||c|}
    \hline
        \textbf{Sources} & Rc/Ra &$\delta^{37}$Cl & $\delta^{81}$Br & Mean Distance by Source\\
        \hline
1 &0.03   &  0.0 &0.0 & 0.01\\
\hline
2 & 0.01  &   0.0 &0.1 & 0.04\\
\hline
3 & 0.02   &  0.0 &0.04 & 0.02\\
\hline\hline
Mean Distance by Dimension&0.02 &    0.0 &    0.05   &  0.02\\
        \hline
    \end{tabular}
    \caption{Distance between the proposed sources computed from the empirical $\theta$ and the sources estimated in \cite{Pint20}.}
\label{tab: error estimated pinti}
\end{table}

%\section{Discussion}
%The robustness test allows to consider that there are no measurement uncertainties. If there is an absolute need to take into consideration measurement uncertainties, the authors propose to generate multiple data sets by disturbing the original points with the measurement uncertainties. However, if the noise is relatively low, the results should be the close to the results which do not take into account measurement uncertainties.

%The ABC Shadow algorithm allows building priors for the interaction parameters of the Hug model. However, the real sources or at least an estimation of the real sources are needed for the parameter estimation. Moreover, the ABC Shadow algorithm is not made to be used with the Gibbs sampling needed for the source detection in multi multidimensional spaces. Hence, a study of the interaction between the ABC Shadow algorithm and the Gibbs sampling is necessary.

\section*{Conclusions and perspectives}
This paper presented the ABC Shadow algorithm in order to determine %\RS{
appropriate parameters for the Hug model. This approach requires expert geologists knowledge. This knowledge is used either for an initial set up of the parameters in order to obtain a first source detection, or for manual detection of these sources. The pre-detected sources are used to estimate the model parameters that are finally plugged in the Hug model detection method in order to get final results.%}
A visual inspection of the visit maps obtained from the detection with new parameters indicates a general improvement of the obtained results. As a perspective, we mention here the continuation of this study while taking into account data uncertainties. This work is in progress.

\section*{Acknowledgments}
This work was performed in the frame of the DEEPSURF project (http://lue.univ-lorraine.fr/fr/impact-deepsurf) at Universit\'e de Lorraine. This work was supported partly by the French PIA project Lorraine Universit\'e d'Excellence, reference ANR-15-IDEX-04-LUE.

\bibliography{bibliographie}

\begin{thebibliography}{}

\bibitem [\protect \citeauthoryear {%
{\scshape\bgroup}Blum{\egroup}%
}{%
{\scshape\bgroup}Blum{\egroup}%
}{%
{\protect \APACyear {2010}}%
}]{%
Blu10}
\APACinsertmetastar {%
Blu10}%
\begin{APACrefauthors}%
{\scshape\bgroup}Blum{\egroup} MG.%
\end{APACrefauthors}%
\unskip\
\newblock
\APACrefYearMonthDay{2010}{}{}.
\newblock
{\BBOQ}\APACrefatitle {Approximate Bayesian computation: a nonparametric
  perspective} {Approximate bayesian computation: a nonparametric
  perspective}.{\BBCQ}
\newblock
\APACjournalVolNumPages{Journal of the American Statistical
  Association}{105}{491}{1178--1187}.
\PrintBackRefs{\CurrentBib}

\bibitem [\protect \citeauthoryear {%
{\scshape\bgroup}Faure{\egroup}%
}{%
{\scshape\bgroup}Faure{\egroup}%
}{%
{\protect \APACyear {1997}}%
}]{%
Fau97}
\APACinsertmetastar {%
Fau97}%
\begin{APACrefauthors}%
{\scshape\bgroup}Faure{\egroup} G.%
\end{APACrefauthors}%
\unskip\
\newblock
\APACrefYear{1997}.
\newblock
\APACrefbtitle {Principles and applications of geochemistry} {Principles and
  applications of geochemistry}\ (\BVOL~625).
\newblock
\APACaddressPublisher{}{Prentice Hall New Jersey, United States,}.
\PrintBackRefs{\CurrentBib}

\bibitem [\protect \citeauthoryear {%
{\scshape\bgroup}Gemmerl{\'e}{\egroup}%
, {\scshape\bgroup}Stoica{\egroup}%
\BCBL {}\ \BBA {} {\scshape\bgroup}Reype{\egroup}%
}{%
{\scshape\bgroup}Gemmerl{\'e}{\egroup}%
, {\scshape\bgroup}Stoica{\egroup}%
\BCBL {}\ \BBA {} {\scshape\bgroup}Reype{\egroup}%
}{%
{\protect \APACyear {2022}}%
}]{%
GemSto22}
\APACinsertmetastar {%
GemSto22}%
\begin{APACrefauthors}%
{\scshape\bgroup}Gemmerl{\'e}{\egroup} D%
, {\scshape\bgroup}Stoica{\egroup} RS%
\BCBL {}\ \BBA {} {\scshape\bgroup}Reype{\egroup} C.%
\end{APACrefauthors}%
\unskip\
\newblock
\APACrefYearMonthDay{2022}{}{}.
\newblock
{\BBOQ}\APACrefatitle {DRlib: a C++ library for marked Gibbs point processes
  simulation and inference} {Drlib: a c++ library for marked gibbs point
  processes simulation and inference}.{\BBCQ}
\newblock
\BIn{} \APACrefbtitle {21st Annual Conference of the International Association
  for Mathematical Geosciences, IAMG 2022.} {21st annual conference of the
  international association for mathematical geosciences, iamg 2022.}
\PrintBackRefs{\CurrentBib}

\bibitem [\protect \citeauthoryear {%
{\scshape\bgroup}Ingebritsen{\egroup}%
, {\scshape\bgroup}Sanford{\egroup}%
\BCBL {}\ \BBA {} {\scshape\bgroup}Neuzil{\egroup}%
}{%
{\scshape\bgroup}Ingebritsen{\egroup}%
, {\scshape\bgroup}Sanford{\egroup}%
\BCBL {}\ \BBA {} {\scshape\bgroup}Neuzil{\egroup}%
}{%
{\protect \APACyear {2006}}%
}]{%
IngSan06}
\APACinsertmetastar {%
IngSan06}%
\begin{APACrefauthors}%
{\scshape\bgroup}Ingebritsen{\egroup} SE%
, {\scshape\bgroup}Sanford{\egroup} WE%
\BCBL {}\ \BBA {} {\scshape\bgroup}Neuzil{\egroup} CE.%
\end{APACrefauthors}%
\unskip\
\newblock
\APACrefYear{2006}.
\newblock
\APACrefbtitle {Groundwater in geologic processes} {Groundwater in geologic
  processes}.
\newblock
\APACaddressPublisher{}{Cambridge University Press}.
\PrintBackRefs{\CurrentBib}

\bibitem [\protect \citeauthoryear {%
{\scshape\bgroup}{\noopsort{Lieshout}{van Lieshout}}{\egroup}%
}{%
{\scshape\bgroup}{\noopsort{Lieshout}{van Lieshout}}{\egroup}%
}{%
{\protect \APACyear {2000}}%
}]{%
Lies00}
\APACinsertmetastar {%
Lies00}%
\begin{APACrefauthors}%
{\scshape\bgroup}{\noopsort{Lieshout}{van Lieshout}}{\egroup} M.%
\end{APACrefauthors}%
\unskip\
\newblock
\APACrefYear{2000}.
\newblock
\APACrefbtitle {Markov Point Processes and their Applications} {Markov point
  processes and their applications}.
\newblock
\APACaddressPublisher{}{Imperial College Press, London}.
\PrintBackRefs{\CurrentBib}

\bibitem [\protect \citeauthoryear {%
{\scshape\bgroup}M{\o}ller{\egroup}%
\ \BBA {} {\scshape\bgroup}Waagepetersen{\egroup}%
}{%
{\scshape\bgroup}M{\o}ller{\egroup}%
\ \BBA {} {\scshape\bgroup}Waagepetersen{\egroup}%
}{%
{\protect \APACyear {2003}}%
}]{%
MollWaag04}
\APACinsertmetastar {%
MollWaag04}%
\begin{APACrefauthors}%
{\scshape\bgroup}M{\o}ller{\egroup} J%
\BCBT {}\ \BBA {} {\scshape\bgroup}Waagepetersen{\egroup} RP.%
\end{APACrefauthors}%
\unskip\
\newblock
\APACrefYear{2003}.
\newblock
\APACrefbtitle {Statistical inference and simulation for spatial point
  processes} {Statistical inference and simulation for spatial point
  processes}.
\newblock
\APACaddressPublisher{}{Chapman and Hall/CRC}.
\PrintBackRefs{\CurrentBib}

\bibitem [\protect \citeauthoryear {%
{\scshape\bgroup}Pinti{\egroup}%
\ \protect \BOthers {.}}{%
{\scshape\bgroup}Pinti{\egroup}%
\ \protect \BOthers {.}}{%
{\protect \APACyear {2020}}%
}]{%
Pint20}
\APACinsertmetastar {%
Pint20}%
\begin{APACrefauthors}%
{\scshape\bgroup}Pinti{\egroup} DL%
, {\scshape\bgroup}{Shouakar-Stash}{\egroup} O%
, {\scshape\bgroup}Castro{\egroup} MC%
, {\scshape\bgroup}{Lopez-Hern{\'a}ndez}{\egroup} A%
, {\scshape\bgroup}Hall{\egroup} CM%
, {\scshape\bgroup}Rocher{\egroup} O%
\BDBL {}{\scshape\bgroup}{Ram{\'\i}rez-Montes}{\egroup} M%
\end{APACrefauthors}%
\unskip\
\newblock
\APACrefYearMonthDay{2020}{}{}.
\newblock
{\BBOQ}\APACrefatitle {The bromine and chlorine isotopic composition of the
  mantle as revealed by deep geothermal fluids} {The bromine and chlorine
  isotopic composition of the mantle as revealed by deep geothermal
  fluids}.{\BBCQ}
\newblock
\APACjournalVolNumPages{Geochimica et Cosmochimica Acta}{}{}{}.
\PrintBackRefs{\CurrentBib}

\bibitem [\protect \citeauthoryear {%
{\scshape\bgroup}Reype{\egroup}%
}{%
{\scshape\bgroup}Reype{\egroup}%
}{%
{\protect \APACyear {2022}}%
}]{%
reype2022probabilistic}
\APACinsertmetastar {%
reype2022probabilistic}%
\begin{APACrefauthors}%
{\scshape\bgroup}Reype{\egroup} C.%
\end{APACrefauthors}%
\unskip\
\newblock
\APACrefYear{2022}.
\unskip\
\newblock
\APACrefbtitle {Probabilistic modelling and Bayesian inference for the analysis
  of geological fluid mixing systems: pattern detection and parameter
  estimation} {Probabilistic modelling and bayesian inference for the analysis
  of geological fluid mixing systems: pattern detection and parameter
  estimation}\ \APACtypeAddressSchool {\BUPhD}{}{}.
\unskip\
\newblock
\APACaddressSchool {}{Universit{\'e} de Lorraine}.
\PrintBackRefs{\CurrentBib}

\bibitem [\protect \citeauthoryear {%
{\scshape\bgroup}Reype{\egroup}%
\ \protect \BOthers {.}}{%
{\scshape\bgroup}Reype{\egroup}%
\ \protect \BOthers {.}}{%
{\protect \APACyear {2020}}%
}]{%
Rey20}
\APACinsertmetastar {%
Rey20}%
\begin{APACrefauthors}%
{\scshape\bgroup}Reype{\egroup} C%
, {\scshape\bgroup}Richard{\egroup} A%
, {\scshape\bgroup}Deaconu{\egroup} M%
\BCBL {}\ \BBA {} {\scshape\bgroup}Stoica{\egroup} RS.%
\end{APACrefauthors}%
\unskip\
\newblock
\APACrefYearMonthDay{2020}{}{}.
\newblock
{\BBOQ}\APACrefatitle {Bayesian statistical analysis of hydrogeochemical data
  using point processes: a new tool for source detection in multicomponent
  fluid mixtures} {Bayesian statistical analysis of hydrogeochemical data using
  point processes: a new tool for source detection in multicomponent fluid
  mixtures}.{\BBCQ}
\newblock
\APACjournalVolNumPages{arXiv preprint arXiv:2009.04132}{}{}{}.
\PrintBackRefs{\CurrentBib}

\bibitem [\protect \citeauthoryear {%
{\scshape\bgroup}Reype{\egroup}%
\ \protect \BOthers {.}}{%
{\scshape\bgroup}Reype{\egroup}%
\ \protect \BOthers {.}}{%
{\protect \APACyear {2022}}%
}]{%
Rey22}
\APACinsertmetastar {%
Rey22}%
\begin{APACrefauthors}%
{\scshape\bgroup}Reype{\egroup} C%
, {\scshape\bgroup}Stoica{\egroup} RS%
, {\scshape\bgroup}Richard{\egroup} A%
\BCBL {}\ \BBA {} {\scshape\bgroup}Deaconu{\egroup} M.%
\end{APACrefauthors}%
\unskip\
\newblock
\APACrefYearMonthDay{2022}{}{}.
\newblock
{\BBOQ}\APACrefatitle {HUG model: an interaction point process for Bayesian
  detection of multiple sources in groundwaters from hydrochemical data} {Hug
  model: an interaction point process for bayesian detection of multiple
  sources in groundwaters from hydrochemical data}.{\BBCQ}
\newblock
\APACjournalVolNumPages{arXiv preprint arXiv:2208.00959}{}{}{}.
\PrintBackRefs{\CurrentBib}

\bibitem [\protect \citeauthoryear {%
{\scshape\bgroup}Stoica{\egroup}%
\ \protect \BOthers {.}}{%
{\scshape\bgroup}Stoica{\egroup}%
\ \protect \BOthers {.}}{%
{\protect \APACyear {2021}}%
}]{%
StoDea21}
\APACinsertmetastar {%
StoDea21}%
\begin{APACrefauthors}%
{\scshape\bgroup}Stoica{\egroup} RS%
, {\scshape\bgroup}Deaconu{\egroup} M%
, {\scshape\bgroup}Philippe{\egroup} A%
\BCBL {}\ \BBA {} {\scshape\bgroup}Hurtado Gil{\egroup}~L.%
\end{APACrefauthors}%
\unskip\
\newblock
\APACrefYearMonthDay{2021}{}{}.
\newblock
{\BBOQ}\APACrefatitle {Shadow Simulated Annealing: A new algorithm for
  approximate {B}ayesian inference of {G}ibbs point processes} {Shadow
  simulated annealing: A new algorithm for approximate {B}ayesian inference of
  {G}ibbs point processes}.{\BBCQ}
\newblock
\APACjournalVolNumPages{Spatial Statistics}{}{}{100505}.
\PrintBackRefs{\CurrentBib}

\bibitem [\protect \citeauthoryear {%
{\scshape\bgroup}Stoica{\egroup}%
\ \protect \BOthers {.}}{%
{\scshape\bgroup}Stoica{\egroup}%
\ \protect \BOthers {.}}{%
{\protect \APACyear {2017}}%
}]{%
StoPhi17}
\APACinsertmetastar {%
StoPhi17}%
\begin{APACrefauthors}%
{\scshape\bgroup}Stoica{\egroup} RS%
, {\scshape\bgroup}Philippe{\egroup} A%
, {\scshape\bgroup}Gregori{\egroup} P%
\BCBL {}\ \BBA {} {\scshape\bgroup}Mateu{\egroup} J.%
\end{APACrefauthors}%
\unskip\
\newblock
\APACrefYearMonthDay{2017}{}{}.
\newblock
{\BBOQ}\APACrefatitle {ABC Shadow algorithm: a tool for statistical analysis of
  spatial patterns} {Abc shadow algorithm: a tool for statistical analysis of
  spatial patterns}.{\BBCQ}
\newblock
\APACjournalVolNumPages{Statistics and computing}{27}{5}{1225--1238}.
\PrintBackRefs{\CurrentBib}

\bibitem [\protect \citeauthoryear {%
{\scshape\bgroup}Yardley{\egroup}%
\ \BBA {} {\scshape\bgroup}Bodnar{\egroup}%
}{%
{\scshape\bgroup}Yardley{\egroup}%
\ \BBA {} {\scshape\bgroup}Bodnar{\egroup}%
}{%
{\protect \APACyear {2014}}%
}]{%
YarBod14}
\APACinsertmetastar {%
YarBod14}%
\begin{APACrefauthors}%
{\scshape\bgroup}Yardley{\egroup} BW%
\BCBT {}\ \BBA {} {\scshape\bgroup}Bodnar{\egroup} RJ.%
\end{APACrefauthors}%
\unskip\
\newblock
\APACrefYearMonthDay{2014}{}{}.
\newblock
{\BBOQ}\APACrefatitle {Fluids in the continental crust} {Fluids in the
  continental crust}.{\BBCQ}
\newblock
\APACjournalVolNumPages{Geochemical Perspectives}{3}{1}{1--2}.
\PrintBackRefs{\CurrentBib}

\end{thebibliography}

\end{document}